\documentstyle[epsfig]{article}
\def\abstract#1{\vskip 7mm 
        \begin{center}{\large Abstract}\par \smallskip
                \begin{minipage}[c]{12cm}
                        \small #1
                \end{minipage}
        \end{center}
}
\def\title#1{\begin{center}{\Large\bf #1}\end{center}}
\def\author#1{\vskip 5mm \begin{center}{#1}\end{center}}
\def\address#1{\begin{center}{\it #1}\end{center}}
\newcommand{\rmd}{{\rm d}}

\newcommand{\mbf}{{\mathbf{f}}}
\newcommand{\bg}{{\mathbf{g}}}

\newcommand{\bq}{{\mathbf{q}}}

\newcommand{\bu}{{\mathbf{u}}}
\newcommand{\bn}{{\mathbf{n}}}
\newcommand{\bv}{{\mathbf{v}}}
\newcommand{\bx}{{\mathbf{x}}}
\newcommand{\bS}{{\mathbf{S}}}

\newcommand{\bX}{{\mathbf{X}}}

\newcommand{\CA}{{\cal A}}
\newcommand{\CB}{{\cal B}}

\newcommand{\CD}{{\cal D}}

\newcommand{\CR}{{\cal R}}
\newcommand{\CW}{{\cal W}}

\newcommand{\average}[1]{\left\langle #1 \right\rangle_\CD}
\newcommand{\laverage}[1]{\left\langle #1 \right\rangle_{\CD_{\rm \bf i}}}
\newcommand{\baverage}[1]{\left\langle #1 \right\rangle_{\CB_R}}
\newcommand{\initial}[1]{{#1_{\rm \bf i}}}
\newcommand{\inI}{{\bf I}}
\newcommand{\inII}{{\bf II}}
\newcommand{\inIII}{{\bf III}}
\makeatletter
\@ifundefined{lesssim}{}{}
\@ifundefined{gtrsim}{}{}
\def\vereq#1#2{\lower3pt\vbox{\baselineskip1.5pt \lineskip1.5pt
\ialign{$\m@th#1\hfill##\hfil$\crcr#2\crcr\sim\crcr}}}
\makeatother

\begin{document}

\title{%
  On Average Properties of Inhomogeneous Cosmologies
  \smallskip \\
  {}
}
\author{%
  Thomas
Buchert\footnote{E-mail:buchert@theorie.physik.uni-muenchen.de}
}
\address{%
  Theoretical Astrophysics Division, 
National Astronomical Observatory \\
2--21--1  Osawa, Mitaka, Tokyo 181--8588, Japan
}
\abstract{
The present talk summarizes the recently progressed state of a systematic
re--evaluation of cosmological models that respect the presence of inhomogeneities. 
Emphasis is given to identifying the basic steps towards an effective (i.e. 
spatially averaged) description of structural evolution, also unfolding the various 
facets of a ``smoothed--out'' cosmology. We shall highlight some results obtained 
within Newtonian cosmology, discuss expansion laws in general relativity within a 
covariant fluid approach, and put forward some promising directions of future research.}

\bigskip

\section{Motivation and Results in Newtonian Cosmology}

\subsection{Bridging the Gap}

Does an inhomogeneous model of the Universe evolve {\it on average} like
a homogeneous solution of Einstein`s or Newton's laws of gravity ? 
This question is not new, at least among relativists who think that the answer
is certainly {\it no}, not only in view of the nonlinearity of the theories mentioned
{}\cite{ellis:relativistic}. 
The problem was and still is the notion of {\it averaging} whose specification and
unambiguous definition turned out to be an endeavor of high magnitude, mainly 
because it is not straightforward to give a unique meaning to the averaging
of tensors, e.g., a given metric of spacetime. This problem seems to lie in the
backyard of relativists who, from time to time, add another effort towards 
a solution of this problem. On the other hand, the community of cosmologists
``should'' locate exactly this problem at the basis of their evolutionary models of
the Universe. Although there are many exceptions (e.g. {}\cite{futamase:backreaction1},
{}\cite{bildhauer:backreaction1}, {}\cite{bildhauer:backreaction2},
{}\cite{futamase:backreaction2}, {}\cite{seljak:hui}, {}\cite{russ:backreaction},
{}\cite{takada:backreaction}; {}\cite{carfora:RG}, 
{}\cite{hosoya:RG}, {}\cite{sota:RG}, a certainly incomplete list), 
most researchers in this field are drawn back to the 
historical development of cosmologies starting with Friedmann, Einstein and de--Sitter
at the beginning of the last century. Despite the drastic changes of our picture
of structures in the Universe on large scales, still, the cosmologist's thinking 
rests on the hegemony of the so--called ``standard model'' (i.e. the family of
FLRW models for homogeneous and isotropic matter distributions).
This standard model, up to the present state of knowledge, explains 
(or better is employed to explain) a wide variety of
orthogonal observations, and it is therefore hard to beat due to its (suggestively)
established status of resistance against observational tests.
Therefore, most discussions in this field are based on the 
vocabulary of the standard model, aiming to constrain its ``cosmological 
parameters'', often on the basis of observations of structure in the regional
Universe that is very different from homogeneous and isotropic.  

Bridging the gap between an involved mathematical problem of general relativity 
and the practical modeling of cosmological dynamics is possible, 
if ambiguities of averagers could be removed, and results  
related to contemporary discussions in observational cosmology. 
In the sequel we shall follow a line of thought that will match this need and 
could intensify work directed to mastering an inhomogeneous spacetime. 

\subsection{Setting the Pace}

Averaging procedures can be defined in a vast variety of ways
(see, e.g., a recent summary by 
Stoeger  {\it et  al.} {}\cite{stoeger:averaging}). A smoothing--out
operator can live on different foliations of spacetime; it can leave an 
averaged field space--dependent, thus operating on a given spatial scale;
it can also smooth out all inhomogeneities. It could involve statistical
ensemble averaging. It can act on scalars, vectors or
tensors; it can smooth matter variables or, most elegantly, the geometry of spacetime 
itself {}\cite{carfora:deformation1}, {}\cite{carfora:deformation2}.
It can average spacetime variables and not merely spatial variables, etc. 
To remove ambiguity here, it
is best to work within the framework of Newtonian cosmology as a first step of 
understanding,
since there the choice of foliation is not a problem and spatial averaging
can be simply put into practice by Euclidean volume integration. As a second step,
we confine ourselves to scalar variables. This allows us to get at least some
corresponding answer in general relativity, since the averaging of scalars
is, for a given foliation, a covariant operation. We therefore propose to look 
at the simplest (mass--conserving) averager as follows.

\subsection{A Newtonian Averager}

Consider any simply--connected spatial domain in Newtonian spacetime.
With $\average{\cdot}$ we denote  spatial averaging in Eulerian space,
e.g.,  for a  spatial tensor  field  $\CA(\bx,t)=\{A_{ij}(\bx,t)\}$ we
simply have the Euclidean volume  integral normalized by the volume of
the domain:
\begin{equation}
\label{eq:average-def}
\average{\CA} (t) = \frac{1}{V(t)} \int_\CD \rmd^3 x \; \CA(\bx,t) \;\;.
\end{equation}
(Here, $\bx$ are non--rotating Eulerian coordinates.) Since, with this averager,
the averaged field is only time--dependent, the space--dependence is only 
implicit by a functional dependence on the domain`s morphology and position.
We shall evolve the domain in time by preserving its mass content. This is 
a natural assumption, if we also want to extend the domain to the whole Universe.

Now,  consider the volume  of an  Eulerian spatial  domain $\CD$  at a
given  time, $V=\int_{\CD}\rmd^3x$,  and follow  the  position vectors
$\bx=\mbf(\bX,t)$ of  all fluid  elements (indexed by  the Lagrangian
coordinates $\bX$) within the domain. Then, the volume elements are deformed
according to  $\rmd^3x=J\rmd^3X$, where $J$ is the  Jacobian determinant of the
transformation  from Eulerian  to Lagrangian  coordinates.   The total
rate of change of the volume  of the same collection of fluid elements
may then be calculated as follows:
\begin{equation}
\label{eq:theta-HD}
\frac{d_t V}{V} = \frac{1}{V} \rmd_t \int_{\initial{\CD}}\rmd^3X\ J 
= \frac{1}{V}\int_{\initial{\CD}}\rmd^3X\  d_t J \\ 
= \frac{1}{V} \int_{\initial{\CD}}\rmd^3X\ \theta J 
= \average{\theta} = :3 H_{\CD},
\end{equation}
where   $d_t$   is  the   total   (Lagrangian)  time--derivative,
$\theta = \nabla \cdot {\bv}=\frac{\dot J}{J}$ the local expansion rate of a given 
velocity model ${\bv}({\bx}, t)$; $\initial{\CD}$ denotes the initial domain
found by mapping back $\CD$ with the help of ${\mbf}^{-1}$ (provided 
${\mbf}^{-1}$ exists), and
$H_{\CD}={\dot  a}_\CD/a_\CD$  naturally defines an effective 
Hubble--parameter  on  the  domain.

\subsection{Commuting Averaging and Evolution}

From the point of view of the standard model of cosmology it makes no difference,
if we smooth the initial inhomogeneities (say, at a time in the matter dominated
epoch) and then evolve the smoothed data with a FLRW solution, or if we evolve
these inhomogeneities until present and then smooth the distribution. 
The conjecture is held that both ways produce the same values for the characteristic
parameters of a FLRW cosmology.

However, averaging, as defined above, and evolving inhomogeneities are non--commuting
operations. To see this we have to
notice  that  the total (Lagrangian) time--derivative does  not
commute with spatial averaging in Eulerian space. For an arbitrary tensor field 
$\CA$ we can readily derive,  with   the  help  of  the  above   definitions,
the  following {\em Commutation Rule} {}\cite{buchert:averaging}:
\begin{equation}
\label{eq:commutation-rule}
\rmd_{t}\average{\CA}-\average{\rmd_{t}\CA}=
\average{\CA\theta}-\average{\theta}\average{\CA}.
\end{equation}

This tells us that exchanging the operators for averaging and time--evolution produces
a source due to the presence of inhomogeneities 
(that we will discuss to consist of positive--definite fluctuations). As an example
consider the expansion rate itself.  Setting ${\CA} = \theta$, we get 

\begin{equation}
\label{eq:commutation-rule_theta}
\rmd_{t}\average{\theta}-\average{\rmd_{t}\theta}=
\average{\theta^2}-\average{\theta}^2 = \average{(\theta - \average{\theta})^2}
\;\ge\;0\;\;,
\end{equation}
i.e., the source is the averaged mean square fluctuation of the expansion rate. 
It vanishes for the case where the local rate equals the global one, which is true for
homogeneous--isotropic matter distributions.

\subsection{Formulating the General Expansion Law of Newtonian Cosmology}

\begin{figure}
\begin{center}
\epsfig{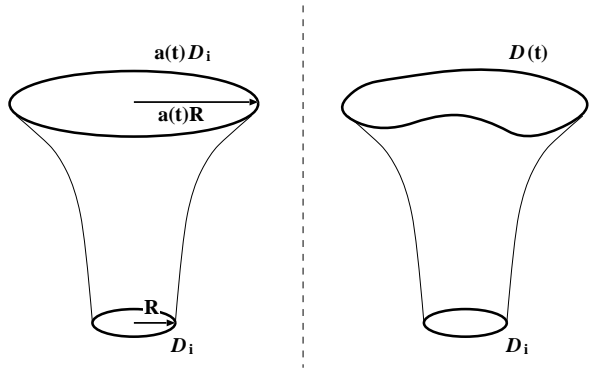}
\end{center}
\caption{\label{fig:jgrg_buchert1}  The  evolution   of  an  initial  domain
$\initial{\CD}$ within the Hubble flow of a FLRW cosmology 
with global scale factor $a(t)$
(left) and for a general inhomogeneous setting (right).}
\end{figure}

As suggested by Eq.~(\ref{eq:theta-HD}) we may measure the effective expansion
of a portion of the Universe with the help of the rate of volume change.
Let us introduce an effective dimensionless scale--factor  $a_\CD$ via the
domain's    volume    $V(t)=|\CD|$     and    the    initial    volume
$\initial{V}=V(\initial{t})=|\initial{\CD}|$ (compare Fig.~\ref{fig:jgrg_buchert1}):
\begin{equation}
\label{eq:def-ad}
a_\CD(t) = \left(\frac{V(t)}{\initial{V}}\right)^{\frac{1}{3}} \;\;,\;\;{\rm
i.e.}\;\;,\;\;H_\CD = \frac{\dot a_\CD}{a_\CD}\;\;.
\end{equation}
For  domains  $\CD$ with  constant  mass  $M_\CD$,  as for  Lagrangian
defined domains, the average mass density evolves as:
\begin{equation}
\label{eq:average-rho}
\average{\varrho} = \frac{\laverage{\varrho(\initial{t})}}{a_\CD^3} 
= \frac{M_\CD}{a_\CD^3 \initial{V}} .
\end{equation}

Looking at the {\em Commutation Rule}, Eq.~(\ref{eq:commutation-rule_theta}),
we appreciate that this rule furnishes a purely kinematical relation provided we give
a velocity model. Dynamics enters by saying how this velocity model is
generated by gravity. In Eq.~(\ref{eq:commutation-rule_theta}) we can express
most of the terms already through the scale--factor and its time--derivatives.
The only unknown is the evolution equation for the local expansion rate. This is
furnished by Raychaudhuri's equation (which employs the Newtonian field equation
$\nabla \cdot {\bg} = \Lambda  - 4\pi G \varrho$ for the gravitational field
strength of dust matter $\bg = \dot\bv$).

Inserting Raychaudhuri's equation, 
\begin{equation}
\label{eq:raychaudhuri}
\dot\theta=\Lambda-4\pi G\rho -\frac{1}{3}\theta^2 + 2 (\omega^2 - 
\sigma^2)\;\;;\;\;\sigma: = \sqrt{\frac{1}{2}\sigma_{ij}\sigma_{ij}}\;\;,\;\;
\omega: = \sqrt{\frac{1}{2}\omega_{ij}\omega_{ 1 
ij}}\;\;, 
\end{equation}
into Eq. (\ref{eq:commutation-rule_theta}), we  find that the
scale--factor    $a_\CD$ {\it in general}    obeys    the  dynamical expansion
law {}\cite{buchert:averaging}:
\begin{equation} 
\label{eq:expansion-law}
3 \frac{{\ddot a}_\CD}{a_\CD} + 4\pi G\average{\varrho} -\Lambda = Q_\CD \;\;,
\end{equation}
with  Newton's gravitational constant  $G$, the  cosmological constant
$\Lambda$, and the source term $Q_\CD$, which we may call ``backreaction  term'',
since it measures the departure from the standard model due to the influence of
inhomogeneities. 

$Q_\CD$  depends on  the kinematical  scalars, the
expansion rate $\theta$,  the rate of shear $\sigma$,  and the rate of
vorticity $\omega$ featuring three positive--definite fluctuation terms:
\begin{equation}
\label{eq:Q-kinematical-scalars}
Q_\CD = \frac{2}{3} \left( \average{\theta^2} - \average{\theta}^2 \right) 
+ 2 \average{\omega^2} - 2\average{\sigma^2} \;\;.
\end{equation}

\subsection{The Quintessence of Inhomogeneities and Anisotropies}

It is easily verified that ${Q_\CD} = 0$ is a necessary and sufficient condition
for having ${a_\CD} = a(t)$, with the global scale--factor $a(t)$ solving the standard 
Friedmann equations. Assuming for the moment that the averaged Universe would be
described by the standard model, then the universal expansion rate would be 
determined by four components: the average mass
density (commonly conceived to have baryonic and non--baryonic ``dark'' components),
the value of the cosmological constant, and the global ``curvature parameter'' $k$,
which arises as an integration constant 
by integrating the second--order Friedmann equation (Eq.~(\ref{eq:expansion-law}) for
$Q_\CD =0$) in order to get the global expansion rate $3 H(t)=3 \frac{\dot a}{a}$.

Upon integrating the actual expansion law,
Eq.~(\ref{eq:expansion-law}),  we instead  obtain
({}\cite{buchert:averaging-hypothesis}    used   a    different   sign
convention for $Q_\CD$):
\begin{equation}
\label{eq:average-friedmann}
3\frac{\dot{a}_\CD^2}{a_\CD^2} + 3\frac{k_\CD}{a_\CD^2 } - 8\pi G \average{\varrho}
- \Lambda = \frac{1}{a_\CD^2} \int_{\initial{t}}^t \rmd t'\ Q_\CD
\frac{\rmd }{\rmd t'} a^2_\CD(t')\;\;,
\end{equation}
where $k_\CD$  enters as  a domain--dependent integration  constant.    
The effective Hubble--parameter $H_\CD=\frac{\dot{a}_\CD}{a_\CD}$ is now
determined by the previous (now domain--dependent) components, but also
by a fifth component due to the presence of inhomogeneities. 

Since there are many unresolved conundrums of the standard model (see, e.g.,
{}\cite{ellis:dahlem}, {}\cite{straumann:lambda}),
cosmologists are searching for a fifth parameter that
may resolve them (see the talk by Prof. Fujii on
``Quintessence'' in the present Proceedings). The use of the letter $Q$
for the ``backreaction'' is coincidential, but it may possibly be  
a working coincidence. 

\subsection{Cosmic Triangle or Cosmic Quartett ?}

The vocabulary of the standard model may be condensed into
the picture of a ``cosmic triangle'' \cite{bahcall:triangle},  consisting of 
three cosmological parameters, which are constructed by normalizing the mass density, 
the cosmological constant, and 
the ``curvature parameter'' by the square of the Hubble--parameter.   

As suggested by Eq.~(\ref{eq:average-friedmann}) we are led to define an additional
dimensionless  ``kinematical  backreaction parameter'' through

\begin{equation}
\label{eq:def-omegaQ}
\Omega_Q^\CD := \frac{1}{3\ a_\CD^2 H_\CD^2} 
\int_{\initial{t}}^t \rmd t'\ Q_\CD \frac{\rmd }{\rmd t'} a^2_\CD(t')\;\;,
\end{equation}
in addition to the common cosmological parameters:
\begin{equation}
\label{eq:omega-local-def}
\Omega_m^\CD := \frac{8\pi G\average{\varrho}}{3H_\CD^2}\;\;, \quad 
\Omega_\Lambda^\CD := \frac{\Lambda}{3 H_\CD^2}\;\;,  \quad  
\Omega_k^\CD := -\frac{k_D}{a_\CD^2 H_\CD^2} \;\;\;.
\end{equation}
However, contrary to the standard model, 
all  $\Omega^\CD$--parameters are now domain--dependent and
transformed into  fluctuating fields on the domain;  for $Q_\CD=0$ the
``cosmic triangle'' is undistorted and the parameters acquire their global
standard     values.   

Comparing these definitions with Eq.~(\ref{eq:average-friedmann}) we have
\begin{equation}
\label{eq:omega-sum}
\Omega_m^\CD \;+\; \Omega_\Lambda^\CD \;+\; \Omega_k^\CD \;+\; \Omega_Q^\CD \;=\; 1 
\;\;,
\end{equation}
i.e., there are four players in the game (or five, respectively, if we split 
$\Omega_m^\CD$ into baryonic matter and non--baryonic ``dark matter''). 

In  Friedmann--Lema\^\i{}tre  cosmologies there  is  by definition  no
backreaction: $\Omega_Q^\CD=0$.  In this case a universe model with
$\Omega_m =1$  conserves the matter parameter. Also,
the initial value of the ``curvature parameter'' $\Omega_k =0$ remains so during
the entire evolution. This changes, if ``backreaction'' is taken into account.
$\Omega_Q^\CD$ itself may act as a ``kinematical
dark matter'', or as a ``kinematical cosmological term'', respectively, 
depending on the relative
strength of shear--, epansion-- and vorticity--fluctuations.

As  we  have learned {}\cite{buchert:bks} and as we shall illustrate below,  
the parameters corresponding to the ``cosmic triangle'' can experience large 
changes, even in situations when the ``backreaction parameter'' is seemingly
negligible, but non--zero.

\subsection{Construction Principle for Evolution Models}

In order to obtain quantitative results on the value and impact of the 
``backreaction parameter'', we have to employ evolution models for the 
inhomogeneities. Their implementation seems not to be straightforward, 
since cosmological evolution models (analytical or N--body simulations) 
are constructed in such a way
that they evolve an initial power spectrum of density-- and velocity--fluctuations on a 
given global background (a solution of the standard Friedmann equations), implying
a vanishing ``backreaction'' by construction.
However, the framework of Newtonian cosmology offers the possibility of making use
of contemporary evolution schemes by assigning sense to a global background within
the general expansion law of an averaged inhomogeneous model. 
We shall now elaborate on this fact by demonstrating the validity of the
following {\it Construction Principle}:

\smallskip
\noindent
{\it The average expansion of a generic inhomogeneous matter distribution on
(topologically) closed Newtonian space sections
is given by the solution of the standard Friedmann equations.}

\smallskip
This principle, although usually not explicitly stated, 
lies at the basis of any evolution model in Newtonian cosmology. It appears to
be very restrictive in light of the general relativistic framework (see 
Subsection 2.5).

In order to proof this principle let us work  with  the  invariants of  the
gradient  of  the  velocity  field\footnote{A  comma  denotes  partial
derivative  with respect  to  Eulerian coordinates  $\partial/\partial
x_i\equiv,i$.}. They are expressible in
terms of kinematical  scalars and can be written  as total divergences
of vector fields, which has been  used and discussed in the context of
perturbation                                                  solutions
({}\cite{buchert:lagrangianthree}, {}\cite{ehlers:newtonian}):
\begin{equation}
\label{eq:v-inv-IandII}
\inI(v_{i,j})  = \nabla \cdot \bv = \theta \;\;\;,\;\;\;
\inII(v_{i,j})
 = \frac{1}{2}\nabla\cdot
\Big(\bv (\nabla\cdot\bv) - (\bv\cdot\nabla)\bv \Big)
= \omega^2 - \sigma^2 + \frac{1}{3} \theta^2 \;\;.  
\end{equation}

The ``backreaction term'' can be entirely expressed in terms of the first and  
second invariants:

\begin{equation}
\label{eq:backreaction-invariants}
Q_\CD = 2 \average{\inII(v_{i,j})} - \frac{2}{3} \average{\inI(v_{i,j})}^2 \;\;.
\end{equation}

Now, let us formally assume that there exists a global, homogeneous and isotropic 
reference model with 
scale--factor $a(t)$. We introduce the following variables with respect to this
reference background.
With  the global  Hubble--parameter  $H=\dot{a}/a$  we  define   comoving
Eulerian  coordinates $\bq  :=\bx/a$ and  peculiar--velocities $\bu
:=\bv-H\bx$ as usual.
Using  the  derivative  $\partial_{q_j}u_i\equiv\partial  u_i/\partial
q_j$ with respect to comoving  coordinates we obtain for the first and
second invariants:
\begin{equation}
\inI(v_{i,j}) =  
3H + \frac{1}{a}   \inI\left(\partial_{q_j}u_i\right) \;\;\;,\;\;\; 
\inII(v_{i,j})
  = 3H^2 + \frac{2H}{a} \inI\left(\partial_{q_j}u_i\right) + 
\frac{1}{a^2} \inII\left(\partial_{q_j}u_i\right) \;\;.
\nonumber
\end{equation}
The ``backreaction term'' remains form--invariant (note: $\frac{1}{a}\partial_{q_j}u_i
= u_{i,j}$):

\begin{equation}
\label{eq:backreaction-invariantscomoving}
Q_\CD = \frac{1}{a^2}\left(2 \average{\inII\left(\partial_{q_j}u_i\right)} - 
\frac{2}{3} \average{\inI\left(\partial_{q_j}u_i\right)}^2 \right)\;\;.
\end{equation}
Thus, all terms corresponding to the background flow cancel in this expression, 
and only inhomogeneities contribute to ``backreaction''.

Using  Eq.~(\ref{eq:v-inv-IandII})   we write  the
``backreaction term'' as  a volume--average over divergences.  Hence, using the
theorem of Gauss we obtain:
\begin{equation}
\label{eq:Q-surface-int}
Q_\CD=\frac{1}{a^2} \left[ 2\ \frac{1}{V_q}\int_{\partial\CD_q} \rmd\bS \cdot
\left(\bu(\nabla_q\cdot\bu)-(\bu\cdot\nabla_q)\bu\right) -\frac{2}{3}\ 
\left(\frac{1}{V_q}\int_{\partial\CD_q}\rmd\bS \cdot \bu \right)^2 \right],
\end{equation}
with the  surface $\partial\CD_q$ bounding the comoving  domain $\CD_q$, the
surface  element  $\rmd\bS$,  and  the  comoving  differential  operator
$\nabla_q$. 

From Eq.~(\ref{eq:Q-surface-int})  we directly obtain  $Q_\CD=0$ for a
domain  with empty  boundary,  e.g.,  for toroidal space sections, or 
periodic  peculiar--velocity fields, respectively.  
In turn, the assumed reference solution is a standard Hubble--flow
and may serve as a global background.
(For more details see {}\cite{buchert:averaging}.)

This establishes the {\it Construction Principle} $\;${\bf q.e.d.}

\subsection{Generalization of the Top--Hat Model}

The {\em Construction Principle} provides room for employing the language
of standard theories of structure formation. As an example we apply the 
general expansion law to domains within a global Hubble--flow.

Assuming an Einstein--de--Sitter background for this example we  subtract  a standard
Friedmann  equation,  $3\frac{{\ddot  a}}{a}+4\pi G\varrho_H=0$,  from
Eq.~(\ref{eq:expansion-law}), with the background density 
$\varrho_H=\frac{3H^2}{8\pi G}$, and
obtain the following differential equation for $a_\CD(t)$ {}\cite{buchert:bks}:
\begin{equation}
\label{eq:evolution-ad}
3\left( \frac{{\ddot a_\CD}}{a_\CD}-\frac{{\ddot a}}{a}\right) +
\frac{3}{2} \left(\frac{{\dot a}}{a}\right)^2 \average{\delta} 
= Q_\CD \;\;,
\end{equation}
with  $\average{\delta}=(\average{\varrho}-\varrho_H)/\varrho_H$
specifying the averaged density  contrast $\delta$ in $\CD$.  For $Q_\CD=0$ and
$\average{\delta}=0$  this  equation   simply  states  that  the  time
evolution of  a domain follows the  global expansion, $a_\CD(t)=a(t)$.
For          $Q_\CD=0$          and
$1+\average{\delta}=\frac{\average{\varrho}  a^3}{\varrho_H  a_\CD^3}$
the evolution of  $a_\CD$ is still of Friedmann type,  but with a mass
different from the background mass.  An important subcase
with $Q_\CD=0$ and $\average{\delta}\ne 0$ is the well--known spherical
top--hat  model {}\cite{peebles:lss}.   In Eq.~(\ref{eq:evolution-ad})
there are  two sources determining  the deviations from  the Friedmann
acceleration, the over/under--density {\em and} the ``backreaction term''.
This shows that, in general, 
the   evolution   of a Newtonian portion of the Universe is triggered by an
over/under--density {\em  and}  velocity--fluctuations. 

\begin{figure}
\begin{center}
\epsfig{figure=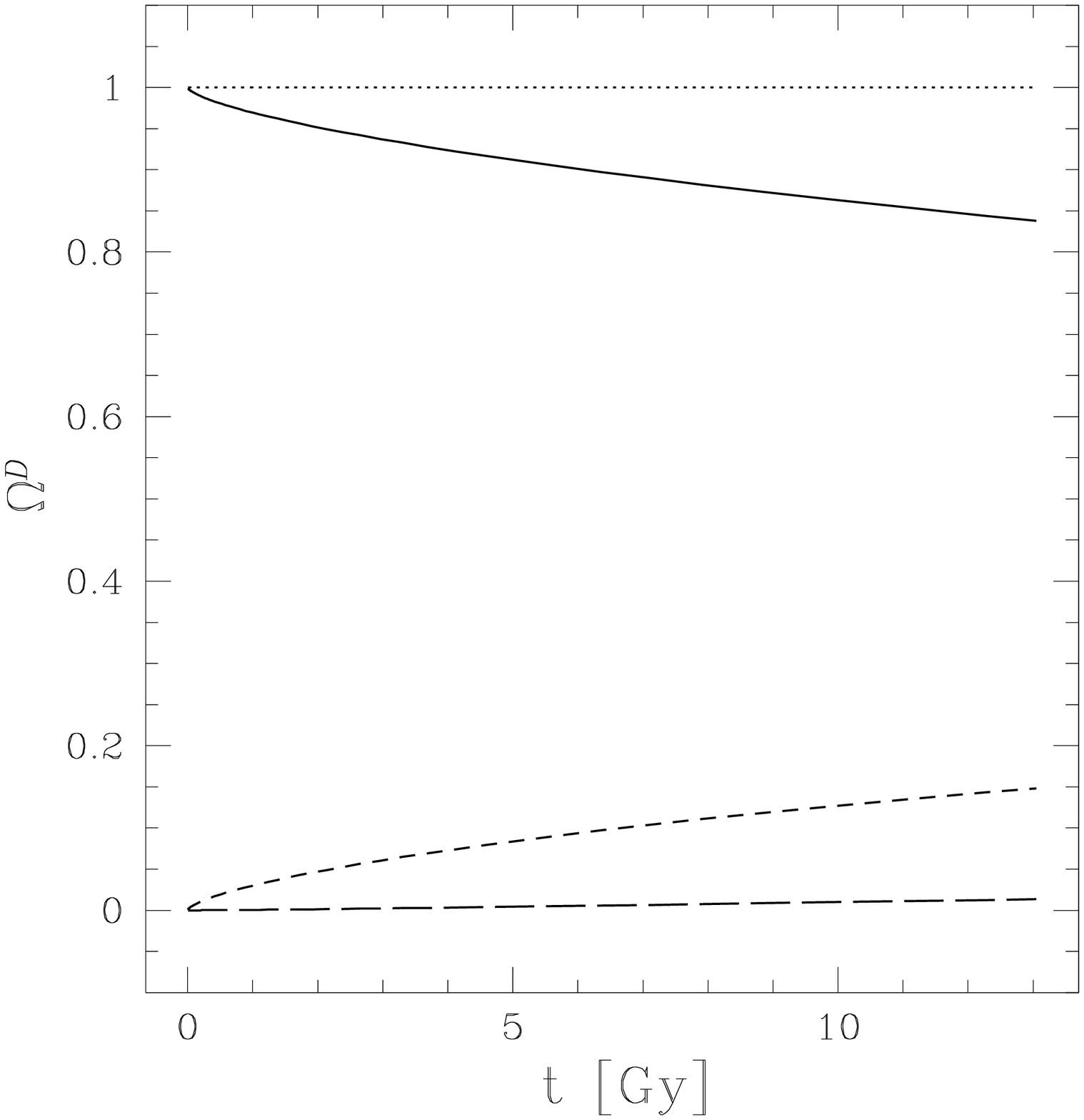,width=7.5cm}
\epsfig{figure=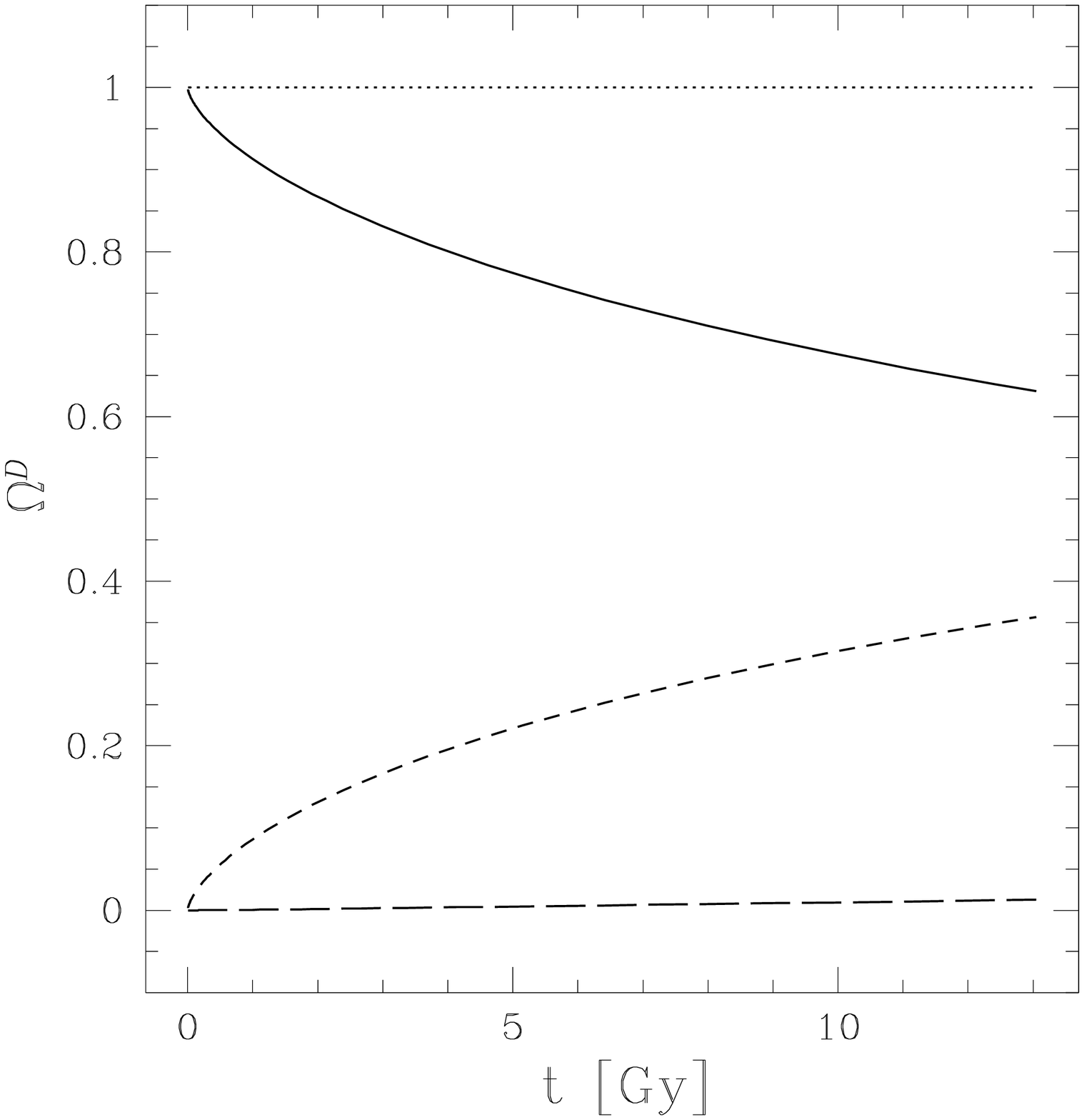,width=7.5cm}
\epsfig{figure=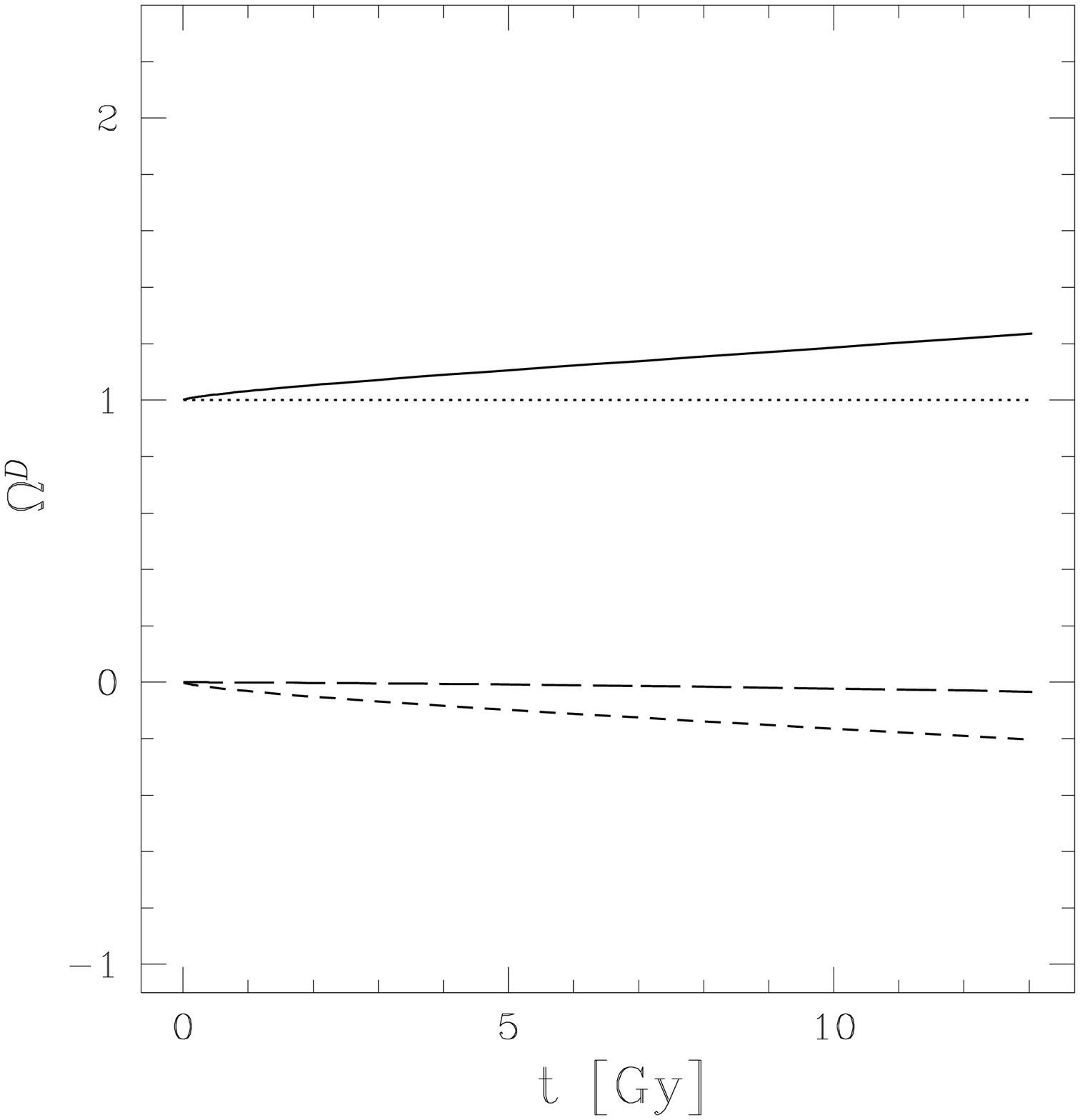,width=7.5cm}
\epsfig{figure=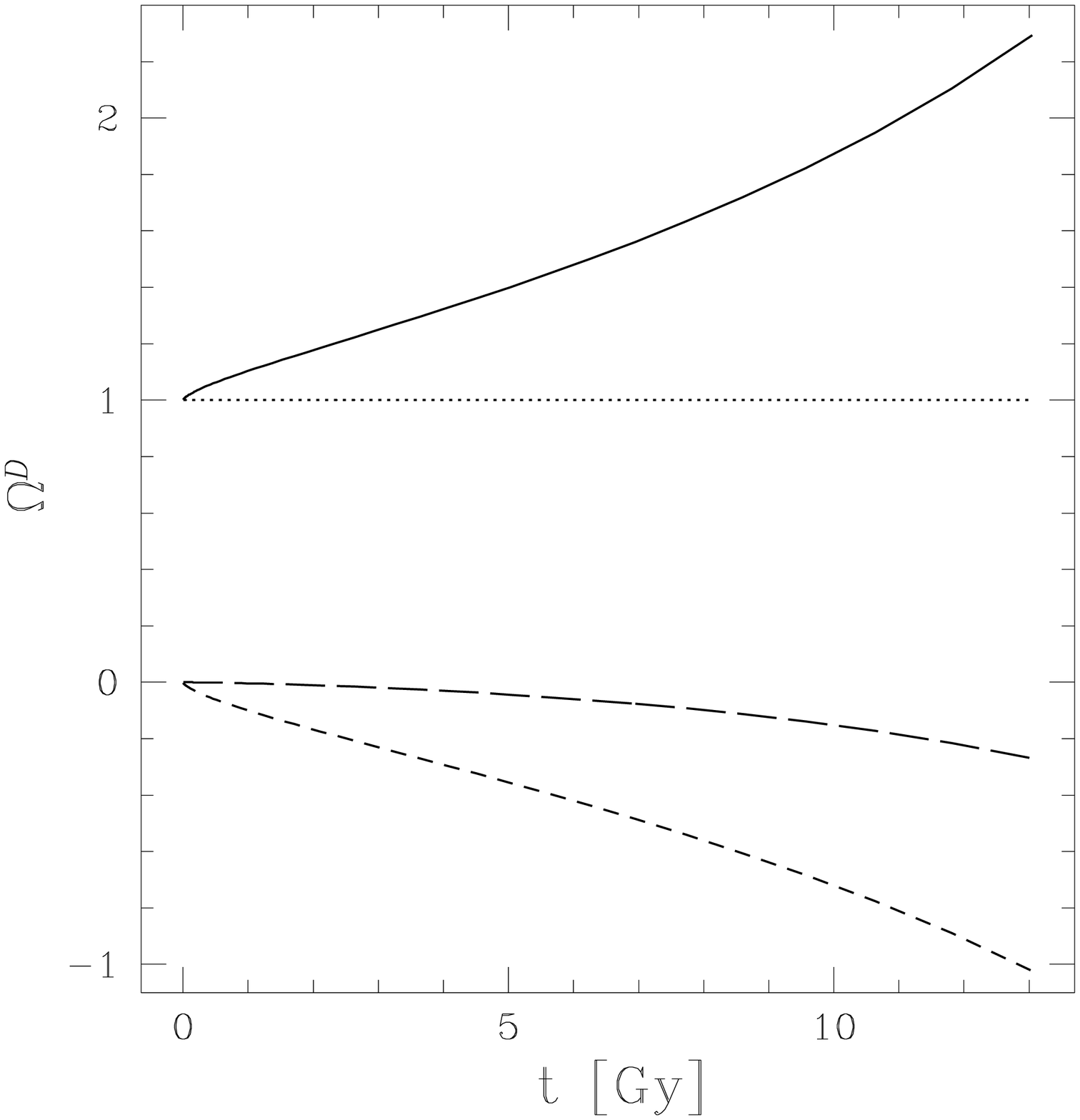,width=7.5cm}
\end{center}
\caption{\label{fig:jgrg_buchert2} The evolution of the ``cosmological
  parameters''  $\Omega_m^\CD$  (solid  line),  $\Omega_k^\CD$  (short
  dashed),   and  $\Omega_Q^\CD$   (long  dashed)   in   an  expanding
  (under--dense) domain with initial radius 0.5Mpc (a scaled radius of
  100Mpc today).         The       dotted       line        is       marking
  $\Omega_m^\CD+\Omega_k^\CD+\Omega_Q^\CD$.
  The upper left  plot  is  for  one--$\sigma$,  the upper right  plot  is  for
  three--$\sigma$  fluctuations (assigned to the expected amplitudes of the initial
  invariants in a Gaussian random field of standard Cold--Dark--Matter fluctuations).  
  The  lower plots show the result for a collapsing domain.
  We see here that the ``backreaction parameter'' is quantitatively more important
  than in an expanding domain -- an interpretation of this result will be offered 
  below: a collapsing domain experiences more drastic changes in shape.
}
\end{figure}

\subsection{An Averaged Lagrangian Perturbation Scheme}

As a second example we apply the {\em Construction Principle} to the
widely used and well--tested Lagrangian perturbation schemes (see: 
{}\cite{sahni:approximation}, {}\cite{ehlers:newtonian} for reviews, and
{}\cite{buchert:lagrangian} for a tutorial).

The trajectory field in the mostly employed ``Zel'dovich approximation''
{}\cite{zeldovich:fragmentation1}, {}\cite{zeldovich:fragmentation2} 
(which can be derived as a subcase of the
first--order Lagrangian perturbation scheme 
{}\cite{buchert:zeldovich}), is given by:
\begin{equation}
\label{eq:zeldovich}
\mbf^Z(\bX,t)=a(t)\Big(\bX+\xi(t)\nabla_0\psi(\bX)\Big) .
\end{equation}
$\psi(\bX)$ is the initial displacement field, $\nabla_0$ the gradient
with   respect  to   Lagrangian   coordinates  and   $\xi (t)$  a   global
time--dependent  function (given for all background models in 
{}\cite{bildhauer:solutions}).

Given this trajectory field we can already approximate the rate of volume 
change by the volume deformation that is caused by the field. For the 
effective scale--factor we obtain {}\cite{buchert:bks}:
\begin{equation}
\label{eq:averageJZ}
(a_\CD^{\rm kin})^3 = 
a^3 \left(1 + \xi\laverage{\initial{\inI}} + 
\xi^2\laverage{\initial{\inII}}
+ \xi^3\laverage{\initial{\inIII}} \right).
\end{equation}

A better estimate is to calculate the ``backreaction term'' from the 
approximation of the velocity field ${\bv^Z} = \dot{\mbf}^Z$ and 
solve the dynamical expansion law, Eq.~(\ref{eq:expansion-law}), for $a_\CD$. 
The ``backreaction term''
separates into its time--evolution given  by $\xi(t)$ and the spatial dependence
on  the  initial  displacement   field  given  by  averages  over  the 
invariants of the displacement gradient 
$\initial{\inI}, \initial{\inII}$, and $\initial{\inIII}$:

\begin{eqnarray}
\label{eq:Q-full-zel}
Q^Z_\CD =  \;\frac{\dot{\xi}^2}{
\left(1 + \xi\laverage{\initial{\inI}} + \xi^2\laverage{\initial{\inII}}
 + \xi^3\laverage{\initial{\inIII}} \right)^{2}} \;\times
\;\;\;\;\;\;\;\;\;\;\;\;\;\;\;\;\;\;\;\;\;\;\;\;\;\;\;\;
\\ \nonumber
 \Big[ 
\left( 2\laverage{\initial{\inII}} - 
  \frac{2}{3}\laverage{\initial{\inI}}^2 \right) +
\xi \left(6\laverage{\initial{\inIII}}-
  \frac{2}{3}\laverage{\initial{\inI}}\laverage{\initial{\inII}}\right)
+ \xi^2 \left(2\laverage{\initial{\inI}}\laverage{\initial{\inIII}}
  -\frac{2}{3}\laverage{\initial{\inII}}^2 \right)  \Big] \;.
\end{eqnarray}
The  numerator of  the first  term is  global and  corresponds  to the
damping   factor that also arises in the Eulerian linear theory
of gravitational instability;   in   an   Einstein--de--Sitter   universe
$\dot{\xi}^2 \propto a^{-1}$.  The denominator  of the first term is the
volume  effect discussed above,  whereas the  second  term  in  brackets features  
the initial ``backreaction'' as a leading term and higher--order terms.

A property of this approximation that renders it 
very useful as an averaged evolution model should be emphasized: 
it is exact for two orthogonal 
symmetry assumptions: for the evolution of plane--symmetric inhomogeneities,
which is a consequence of the known properties of the first--order Lagrangian
approximation {}\cite{buchert:class}, and for the evolution of spherically 
symmetric inhomogeneities, as will be shown below.

Employing the averaged Lagrangian scheme we can quantify the impact of 
backreaction on the domain--dependent ``cosmological parameters''. For details 
the reader may look at a recent work
{}\cite{buchert:bks}. Here, I would like to mention the 
remarkable result that for  an accelerating,  i.e., under--dense
region,    $\Omega_Q^\CD$    may   be numerically   negligible  as seen in
(Fig.~\ref{fig:jgrg_buchert2}, upper plots),  but  dramatic   changes  in  the  other
parameters are observed. For a  collapsing domain with a present--day radius
of 100Mpc  the mass parameter  of the domain  may even differ  by more
than      100\%      from      the     global      mass      parameter
(Fig.~\ref{fig:jgrg_buchert2}, lower plots).

\subsection{Newton's Iron Spheres}

Looking at the general expansion law, Eq.~(\ref{eq:expansion-law}), the careful
reader may object that Friedmann's equations also hold, if we study the motion
of a spherically symmetric domain, {\em} although the matter distribution inside
the sphere may be inhomogeneous. This fact is known as Newton's ``Iron Sphere
Theorem''. Let us show that the general expansion law respects this theorem.

We note that, for a spherically symmetric distribution of matter within the domain, the
invariants of the velocity gradient, averaged over a ball ${\CB_R}$, 
obey relations resulting in {}\cite{buchert:bks}:
\begin{equation}
\baverage{{\inII}(v_{i,j})} = \frac{1}{3} \baverage{{\inI}(v_{i,j})}^2 ,
\quad {\rm and} \quad
\baverage{{\inIII}(v_{i,j})} = \frac{1}{27} \baverage{{\inI}(v_{i,j})}^3 .
\end{equation}
Inserting this into the ``backreaction term'', Eq.~(\ref{eq:backreaction-invariants}), 
shows that 
$Q^{\rm    spherical}_{\CB_R}=0$ in accordance with Newton's theorem.
The inhomogeneous model discussed in the last subsection also has this property:
inserting the above expressions into Eq.~(\ref{eq:Q-full-zel}) (using the 
proportionality of displacement gradient and initial velocity gradient), we also
obtain   $Q^Z_{\CB_R}=0=Q^{\rm    spherical}_{\CB_R}$.

\subsection{Averaging and the Evolution of Form}

The expansion law, Eq.~(\ref{eq:expansion-law}), is built on the rate of change
of a simple morphological quantity, the volume content of a domain. Although
functionally it depends on other morphological characteristics of a domain, it
does not explicitly provide information on their evolution. An evolution equation
for the ``backreaction term'' is missing. This fact touches
on the problem of closing the hierarchy of dynamical evolution equations
considered as a set of coupled ordinary differential equations in Lagrangian space.
The problem of closing such a hierarchy of equations is often considered in the 
literature and various closure conditions are formulated 
(e.g., {}\cite{bertschinger:hui}), one of them being the ``Silent Universe Model''
in general relativity, which assumes a vanishing magnetic part of the Weyl tensor
{}\cite{bruni:silent}. Averaging such a hierarchy would result in evolution 
equations for the ``backreaction term'' and would, with some local closure 
condition, also close the system of averaged equations. 
Here, we will not pursue this problem further, but instead begin to develop the 
morphological point of view  that eventually implies an alternative
proposition of a morphological closure condition.  

Let us focus our attention on the boundary of the spatial domain $\CD$. 
A priori, the location of this boundary in space enjoys some freedom which we may
constrain by saying that the boundary coincides with a velocity front of the fluid
(hereby restricting attention to irrotational flows).
This way we employ the Legendrian point of view of velocity fronts that is dual
to the Lagrangian one of fluid trajectories. 
Let $S(x,y,z,t) = s(t)$ define a velocity front at Newtonian time $t$, ${\bv}=
\nabla S$. Below, we shall need that
the three principal scalar invariants of the velocity gradient 
$v_{i,j} =: S_{,ij}$
can be transformed into divergences of vector fields as 
written explicitly in Eqs.~(\ref{eq:v-inv-IandII},\ref{eq:v-inv-III}).

Defining the unit normal vector $\bn$ on the front, 
${\bn} = \pm {\nabla S \over |\nabla S |}$ (the sign depends on the direction
of its motion),
the average expansion rate can be written as a flux integral using Gauss' theorem:

\begin{equation}
\average{\theta} = \frac{1}{V}\int_{\CD} d^3 x \,\nabla\cdot \bv = \frac{1}{V}
\int_{\partial\CD} \,{\bf {dS}}\cdot\bv \;\;,
\end{equation}
and, with the surface element $d\sigma$, ${\bf {dS}} = {\bn} d\sigma$,  
we obtain the intuitive result that the 
average expansion rate is related to another morphological quantity of the domain,
the total area of the enclosing surface:

\begin{equation}
\average{\theta} = \frac{1}{V}\int_{\partial\CD}\,d\sigma |\nabla S | \;\;.
\end{equation}
Inserting the velocity potential also into the other invariants,
Eq.~({\ref{eq:v-inv-IandII}) and

\begin{equation}
\label{eq:v-inv-III}
\inIII(v_{i,j}) 
 = \frac{1}{3}\nabla\cdot\left( \frac{1}{2}\nabla\cdot
\Big( \bv(\nabla\cdot\bv) - (\bv\cdot\nabla)\bv \Big) \bv - 
\Big(\bv(\nabla\cdot\bv) - (\bv\cdot\nabla)\bv \Big)\cdot\nabla\bv \right) \;\;, 
\end{equation}
and performing the spatial average, we obtain {}\cite{buchert:morphology}:

\begin{equation}
\average{\inII}=\frac{1}{V} \int_{\CD} d^3 x \, {\inII} =\int_{\partial\CD}\, 
d\sigma |\nabla S |^2 \,{\bf H}
\;;\;\;\;
\average{\inIII} = \frac{1}{V} 
\int_\CD d^3 x \, {\inIII} =\int_{\partial\CD} d\sigma |\nabla S |^3 \,{\bf K}\;\;,
\end{equation}
where $\bf H$ is the mean curvature and $\bf K$ 
the Gaussian curvature of the $2-$surface bounding the domain.
$|\nabla S | = \frac{ds}{dt}$ equals $1$, if the instrinsic arc--length $s$ of the
trajectories is used instead of the extrinsic Newtonian time $t$.
The averaged invariants comprise, together with the volume a complete set of
morphological characteristics related to the
{\em Minkowski functionals of a body}:

\begin{equation}
\CW_0 (s): = \int_{\CD} d^3 x = V \;\;;\;\;
\CW_1 (s): = \frac{1}{3}\int_{\partial\CD} d\sigma \;\;;\;\; 
\CW_2 (s): = \frac{1}{3}\int_{\partial\CD} d\sigma \;{\bf H}\;\;;\;\;
\CW_3 (s): = \frac{1}{3}\int_{\partial\CD} d\sigma \;{\bf K}=\frac{4\pi}{3}\chi\;\;.
\end{equation}
The Euler--characteristic $\chi$ determines the topology of the domain and 
is assumed to be an integral of motion ($\chi = 1$), 
if the domain should remain simply--connected (a morphological closure condition).

Thus, we have gained a morphological interpretation of the ``backreaction term'': it 
can be entirely expressed through three of the four Minkowski functionals:

\begin{equation}
\label{eq:backreaction-minkowski}
Q_\CD (s)= 6  \left(\frac{\CW_2}{\CW_0} - \frac{\CW_1^2}{\CW_0^2}\right) \;\;.
\end{equation}

The $\CW_{\alpha}\;;\;\alpha=0,1,2,3$ have been introduced as 
``Minkowski functionals'' 
into cosmology by Mecke {\it et al.} ({}\cite{mecke:mf}) 
in order to statistically assess morphological properties of cosmic structure.
Minkowski functionals proved to be useful tools to also incorporate
information from higher--order correlations, e.g., in the distribution of galaxies,
galaxy clusters, density fields or cosmic microwave background temperature maps
({}\cite{kerscher:mf}, {}\cite{kerscher:cl}, {}\cite{schmalzing:mf},
{}\cite{schmalzing:cmb}; see the review by Kerscher 
{}\cite{kerscher:review} and ref. therein). Related to the
morphology of individual domains is the study of 
building blocks of large--scale cosmic structure 
{}\cite{sahni:web}, {}\cite{schmalzing:web}.

For a ball with radius $R$ we have for the Minkowski functionals:
\begin{equation}
\CW^{\CB_R}_0 (s): = \frac{4\pi}{3}R^3\;\;;\;\;
\CW^{\CB_R}_1 (s): = \frac{4\pi}{3}R^2 \;\;;\;\; 
\CW^{\CB_R}_2 (s): = \frac{4\pi}{3}R\;\;;\;\;
\CW^{\CB_R}_3 (s): = \frac{4\pi}{3}\;\;.
\end{equation}
Inserting these expressions into the ``backreaction term'',
Eq.~(\ref{eq:backreaction-minkowski}), shows that $Q^{\CB_R}_\CD (s) = 0$, 
and we have confirmed
Newton`s ``Iron Sphere Theorem'' once more. Moreover, we can understand now that 
the ``backreaction term'' encodes the deviations of the domain`s morphology from 
that of a ball, a fact which we shall illustrate now with the 
help of Steiner`s formula of integral geometry (see also {}\cite{mecke:mf}).

Let $d\sigma_0$ be the surface element on the
unit sphere, then (according to the Gaussian map) 
$d\sigma = R_1 R_2 d\sigma_0$ is the surface element of a $2-$surface
with radii of curvature $R_1$ and $R_2$. Moving the surface a distance
$\varepsilon$ along its normal we get for the surface element of the parallel 
velocity front:

\begin{equation} 
\label{eq:steiner}
d\sigma_{\varepsilon} = (R_1 + \varepsilon) (R_2 + \varepsilon) d\sigma_0 = 
\frac{R_1 R_2 + \varepsilon (R_1 + R_2 ) + \varepsilon^2 }{R_1 R_2}d\sigma = 
(1 + \varepsilon 2 {\bf H} + \varepsilon^2 {\bf K}) d\sigma \;\;,
\end{equation}  
where 
\begin{equation}
{\bf H} = \frac{1}{2}\left(\frac{1}{R_1} + \frac{1}{R_2}\right)\;\;\;,
\;\;\; {\bf K} = \frac{1}{R_1 R_2} \;\;,
\end{equation}
are the mean curvature and Gaussian curvature of the front as before.

Integrating Eq.~(\ref{eq:steiner}) over the whole front we arrive at a relation 
between the total surface area $A$ of the front and $A_{\varepsilon}$ of its
parallel front. The gain in volume may then be expressed by an integral of
the resulting relation with respect
to $\varepsilon$ (which is known as {\em Steiner`s formula} defining the
Minkowski functionals of a (convex) body in three spatial dimensions):

\begin{equation}
V_{\varepsilon} =  V + \int_0^{\varepsilon}\, d\varepsilon' A_{\varepsilon'} =
V + \varepsilon A + \varepsilon^2 \int_{\partial\CD}\,d\sigma\;{\bf H} + 
\varepsilon^3  \int_{\partial\CD}\,d\sigma\;{\bf K} \;\;.
\end{equation}

\newpage

\section{Average Models in General Relativity}

\subsection{An Averager for Scalars}

The key to reducing ambiguity of choosing an averager in general relativity is to
confine ourselves to scalars, since spatially 
averaging a scalar  field $\Psi$ is a covariant operation.
However, we have to make sure that we understand the word ``spatial'' physically.
Geometrically, we shall introduce a foliation of spacetime, but we shall encounter
ambiguity in choosing it.
Suppose for the moment that we insist on {\em spatial} averaging and 
consider for all what follows a foliation of spacetime as given.
The simplest averager is then immediately written down as follows:
\begin{equation}
\label{eq:average-GR}
\average{\Psi (X^i , t)}: = 
\frac{1}{V}\int_\CD J d^3 X \;\;\Psi (X^i , t) \;\;\;;\;\;\;V = \int_\CD J d^3 X\;\;\;,
\end{equation}
with  $J:=\sqrt{\det(g_{ij})}$, where  $g_{ij}$ is  the metric  of the
spatial  hypersurfaces, and  $X^i$ are  coordinates in the hypersurfaces,
conveniently chosen such that they are constant
along  flow lines. 

\subsection{Foliating Spacetime: Covariant Fluid Approach}

We wish to conserve mass inside a spatial domain of spacetime. Therefore,
let us first consider the (conserved) restmass flux 
vector\footnote{Greek indices run through
$0 ... 3$, while latin indices run through $1 ... 3$ as before; 
summation over repeated indices is understood. A semicolon will denote 
covariant derivative with respect to the 4--metric with signature $(-,+,+,+)$; 
the units are such that $c=1$.}

\begin{equation}
M^{\mu}: = \varrho u^{\mu}\;\;\;;\;\;\;M^{\mu}_{\;\,;\mu}=0\;\;\;
;\;\;\;\varrho > 0\;\;\;,
\end{equation}
where $\varrho$ is the restmass density and the flow lines are integral
curves of the 4--velocity $u^{\mu}$. 
We shall confine ourselves to irrotational perfect fluids 
with energy density $\varepsilon$ and pressure $p$, which allows us to simplify
the splitting of spacetime, e.g. the spatial hypersurfaces can  be chosen 
flow--orthogonal in this case, the unit normal on the hypersurfaces coincides with
the 4--velocity and the shift vector vanishes.   
Irrotationality guarantees the existence of a scalar function $S$, such that

\begin{equation}
u^{\mu} =: \frac{-\partial^{\mu}S}{h}\;\;\;,
\end{equation}
In general, we identify the magnitude $h$ with the 
``injection energy per fluid element and unit
restmass'',
 
\begin{equation}
h := \frac{\varepsilon + p}{\varrho}\;\;,
\end{equation}
which is related to the relativistic enthalpy 
$\eta: = \frac{\varepsilon + p}{n}$ 
by $h = \eta/m$ with $m$ the unit restmass of a fluid element,
and $n$ the baryon density.
Note that $d\varepsilon = h d\varrho$.
For a barotropic fluid we can easily see that $\varepsilon$ is a function of the
restmass density only and, hence, $h$ is a function of $\varrho$.
$h$ is identical to $1$ in the case of dust.
The magnitude $h$ normalizes the 4--gradient 
$\partial^{\mu}S$ so that $u^{\mu}u_{\mu} = -1$,

\begin{equation}
h = \sqrt{-\partial^{\alpha}S \partial_{\alpha}S} = 
u^{\mu} \partial_{\mu}S = 
{\dot S}\;>\;0\;\;.
\end{equation}

\begin{figure}
\begin{center}
\epsfig{figure=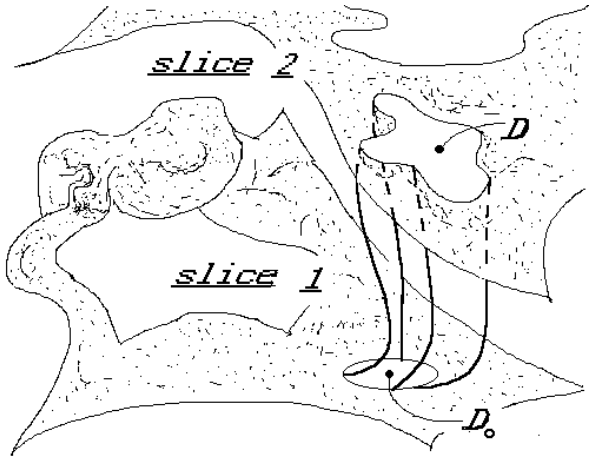,width=8cm} 
\end{center}
\caption{\label{fig:jgrg_buchert3}  The  evolution   of  an  initial  domain
${\CD}_{\bf o}$ within a simply--connected tube of spacetime. The figure
visualizes the restrictions implied by the topological requirement of 
simple connectivity throughout the evolution of the domain. Note that 
the boundary of the domain may break in the coarse of structure 
formation (occurence of a Legendrian singularity), which may induce topology changes.}
\end{figure}

The overdot stands for the material derivative operator along the flow lines
of any tensor field $\cal F$ as defined covariantly by
$\dot{\cal F}: = u^{\mu}{\cal F}_{\,\;;\mu}$.  
It reduces in the case of a scalar function in a flow--orthogonal slicing 
(which we want to envisage)
to the partial time--derivative multiplied by $\frac{1}{N}$, where $N$ is the
lapse function (for more details see: {}\cite{buchert:grgfluid}).

It can be shown that $S$ is spatially homogeneous. Thus, $S(t)$ and $h(X^i ,t)$
play the role of ``phase'' and ``amplitude'' of the fluid's wave fronts.
Foliating the spacetime into hypersurfaces $S(t)=const.$ is a gauge condition
that is naturally adapted to the fluid itself and, therefore, this foliation 
can be given a covariant meaning ({}\cite{bruni:covariant1}, 
{}\cite{bruni:covariant2} and ref. therein). 

\subsection{A Commutation Rule in General Relativity}

The rate of change of the volume $V(t)$ in the hypersurfaces $S(t)=const.$
is evaluated by taking the partial time--derivative of the volume and dividing by
the volume.
Since $\partial_t$ and $d^3 X$ commute (but not ${d\over d\tau}:=\frac{\partial_t}{N}$ 
and $d^3 X$ !) we obtain:
\begin{equation}
\frac{\partial_t V}{V} = \frac{1}{V}
\int_{\CD} d^3 X \partial_t  J = \frac{1}{V}
\int_{\CD} d^3 X N {\dot J}  = \frac{1}{V}
\int_{\CD} d^3 X N\theta J = \average{N\theta}\;\;.
\end{equation}
Introducing the scaled (t--)expansion ${\tilde\theta}:=N\theta$ we define
an effective (t--)Hubble function in the hypersurfaces by
\begin{equation}
\langle\tilde\theta\rangle_{\CD}  = \frac{\partial_t V}{V} = 3 
\frac{\partial_t a_{\CD}}{a_{\CD}} =: 3 {\tilde H}_{\CD}\;\;.
\end{equation}

With the help of these definitions we readily derive the {\em Commutation Rule}:

\begin{equation}
\partial_t \average{\Psi} - \average{\partial_t \Psi}
= \langle{\Psi\tilde\theta}\rangle_{\CD} -
\average{\Psi}\langle{\tilde\theta}\rangle_{\CD}
\;\;.
\end{equation}
The lapse--weighted quantities have to be introduced only in the case of 
non--vanishing pressure, since the pressure gradient induces deviations from a 
geodesic flow implying an inhomogeneous lapse function. 
In the much simpler case of a dust matter model,
the lapse function is homogeneous and can be chosen equal to $1$, 
and the covariant fluid gauge
is identical to the comoving and synchronous gauge, which makes the 
correspondence to the Newtonian investigation most transparent. 

In the sequel we shall, for simplicity, discuss the case of a pressure--less
fluid only and resume the discussion of the more general case thereafter. 

\subsection{Expansion Law of General Relativity: Dust Models}

Employing the averaging procedure outlined above and following the line of thought
of the Newtonian investigation in Section 1, we derive the general expansion law
for dust matter in the synchronous and comoving gauge. The result will be covariant
with respect to this foliation.

I here give the result derived in {}\cite{buchert:onaverage}:
the spatially  averaged equations  for  the scale--factor
$a_\CD$, respecting mass conservation, read:

{\em averaged Raychaudhuri equation:}
\begin{equation}
\label{eq:expansion-law-GR}
3\frac{{\ddot a}_\CD}{a_\CD} + 4\pi G 
\frac{M_\CD}{\initial{V}a_\CD^3} - \Lambda = {Q}_\CD \;\;;
\end{equation}

{\em averaged Hamiltonian constraint:}
\begin{equation}
\label{eq:hamiltonconstraint}
\left( \frac{{\dot a}_\CD}{a_\CD}\right)^2 - \frac{8\pi G}{3}
\frac{M_\CD}{\initial{V}a_\CD^3} + \frac{\average{\CR}}{6} 
- \frac{\Lambda}{3} = -\frac{Q_\CD}{6} \;\;,
\end{equation}
where   the  mass   $M_\CD$,   the  averaged   spatial  Ricci   scalar
$\average{\CR}$   and   the    ``backreaction   term''   $Q_\CD$   are
domain--depen\-dent and, except the mass, time--depen\-dent functions.
The backreaction source term is given by
\begin{equation}
\label{eq:Q-GR} 
Q_\CD : = 2 \average{\inII} - \frac{2}{3}\average{\inI}^2 =
\frac{2}{3}\average{\left(\theta - \average{\theta}\right)^2 } - 
2\average{\sigma^2} \;\;.
\end{equation}
Here,  $\inI$  and $\inII$  denote  the  invariants  of the  extrinsic
curvature  tensor that  correspond  to the  kinematical invariants  we
employed earlier.  The  same  expression  (except  for the  vorticity)  as  in
Eq.~(\ref{eq:Q-kinematical-scalars}) follows  by introducing the split
of the  extrinsic curvature into  the kinematical variables  shear and
expansion (second equality above).

We  appreciate an  intimate correspondence  of the  GR  equations with
their   Newtonian   counterparts   (Eq.~(\ref{eq:expansion-law})   and
Eq.~(\ref{eq:average-friedmann})).  The  first  equation  is  formally
identical  to  the  Newtonian   one,  while  the  second  delivers  an
additional   relation   between  the   averaged   curvature  and   the
``backreaction  term'' that  has no  Newtonian analogue.  This  implies an
important  difference   that  becomes  manifest  by   looking  at  the
time--derivative     of     Eq.~(\ref{eq:hamiltonconstraint}).     The
integrability  condition   that  this  time--derivative   agrees  with
Eq.~(\ref{eq:expansion-law-GR})  is non--trivial in  the GR  context and
reads:
\begin{equation}
\label{eq:integrability-GR}
\partial_t Q_\CD + 6 \frac{{\dot a}_\CD}{a_\CD} Q_\CD +  
\partial_t \average{\CR}
+ 2 \frac{{\dot a}_\CD}{a_\CD} \average{\CR} = 0 \;\;.
\end{equation}
The correspondence  between the Newtonian  $k_\CD$--pa\-ra\-meter and
the averaged spatial Ricci curvature is  more involved in the presence of a
``backreaction term'':
\begin{equation}
\label{eq:integrability-integral-GR}
\frac{k_\CD}{a_\CD^2} - \frac{1}{3} a_\CD^2 \int_{t_i}^t \,dt' \;
Q_\CD\; \frac{d}{dt'} a^2_\CD(t')
= \frac{1}{6}\left(\langle {\cal R} \rangle_\CD + Q_\CD\right) . 
\end{equation}
The  time--derivative  of Eq.~(\ref{eq:integrability-integral-GR})  is
equivalent        to         the        integrability        condition
Eq.~(\ref{eq:integrability-GR}).        Eq.~(\ref{eq:integrability-GR})
shows  that  averaged curvature  and ``backreaction  term'' are  directly
coupled  unlike in  the  Newtonian case,  where the  domain--dependent
$k_\CD$--parameter is  fixed by the initial  conditions.

For dust models in general relativity we may therefore introduce 
dimensionless average characteristics that are slightly different 
from those in Newtonian cosmology, the difference being that the ``backreaction
parameter'' is now directly expressible in terms of the ``backreaction source term'',
and not just as an integral expression:

\begin{equation}
\Omega_m^{\CD} : = \frac{8\pi G M_{\CD}}{3 V_{\initial{\CD}}a_{\CD}^3 
H_{\CD}^2 } \;\;;\;\;
\Omega_{\Lambda}^{\CD} := \frac{\Lambda}{3 H_{\CD}^2 }\;\;;\;\;
\Omega_{k}^{\CD} := - \frac{\average{\cal R}}{6 H_{\CD}^2 }\;\;;\;\;
\Omega_{Q}^{\CD} := - \frac{{\cal Q}_{\CD}}{6 H_{\CD}^2 } \;\;,
\end{equation}
which also obey Eq.~(\ref{eq:omega-sum}).

The evolution of these parameters is intimately related, unlike the 
situation in the standard model,  as may be illustrated by the following equation
{}\cite{buchert:onaverage}:

\begin{equation}
{\dot\Omega}^{\CD}_{Q} + 6H_{\CD} \Omega_{Q}^{\CD}
(1- \Omega_{k}^{\CD} - \Omega_{Q}^{\CD}) + {\dot\Omega}_{k}^{\CD} + 
2H_{\CD}\Omega_{k}^{\CD}(1- \Omega_{k}^{\CD} - \Omega_{Q}^{\CD}) 
- 3H_{\CD}(1- \Omega_{\Lambda}^{\CD} - \Omega_{k}^{\CD} - 
\Omega_{Q}^{\CD})(\Omega_{k}^{\CD} + \Omega_{Q}^{\CD}) \;=\; 0\;.
\end{equation}
It is, e.g.,  a good excercise to show with the help of this equation that an inhomogeneous
model (including ``backreaction''), in which the matter parameter stays $1$ like in
the standard Einstein--de Sitter cosmos, does not exist.

\newpage

\subsection{Liberation from Strict Meter}

Let us now hold in for a moment and sort out what the Newtonian and the relativistic
expansion laws for dust matter distinguish. Their close correspondence bears the
temptation of overlooking a crucial conceptual challenge for the modeling of
inhomogeneities in general relativity.

Given the Newtonian average model, Eq.~(\ref{eq:expansion-law}), 
and its quantitative consequences
(elaborated in detail in {}\cite{buchert:bks}), we can draw the conclusion
that we only have to consider spatial scales that are large enough to have
a negligible influence from the ``backreaction term''. This term, which brought the
higher voltage of having mastered a generic inhomogeneous Newtonian cosmology,
shows no global relevance, and it seems that we are drawn back to the previous state of 
low visibility of the standard cosmological models.  

The key--reason for this outcome is the validity of the {\em Construction Principle}.
We force a globally vanishing ``backreaction'' by the way we construct 
inhomogeneous evolution models. (Remember that the majority of cosmological N--body
simulations and analytical approximations are Newtonian 
and assume periodic boundary conditions for inhomogeneities on a FLRW background.) 
The apparent global irrelevance of ``backreaction'' should not be a surprise, 
since our evaluations have taken place within a periodic cube into which the 
universe model was forced:
the scale of this cube introduces a ``strict meter'' governed by the scale--factor
of a standard FLRW universe, the action of fluctuations of inhomogeneities is
confined to the interior of this cube. 

The relativistic average model not only has a somewhat richer tone by linking the
averaged spatial Ricci curvature to the ``backreaction term''
as a result of the Hamiltonian constraint, it also places an 
equal stress on both of them in the following sense.  
Suppose we let the self--gravitating fluid evolve freely and link its structure to the
curvature of space sections through Einstein's equations. Let now the inhomogeneities 
evolve out of an almost
homogeneous and isotropic state on almost flat space sections. 
Naturally, the inhomogeneities grow in the coarse of structure formation. 
The average Ricci curvature of the space sections, however, is dictated by the 
fluid structure unlike in the standard model which would suggest an evolution 
inversely proportional to the ``strict meter'' $a(t)$. In contrast, the value
of the averaged curvature is an outcome 
rather than a model assumption. Thus, a genuine property of the relativistic model, 
also globally, is a change of the average curvature that does not follow a predesigned 
global law; a priori it is not constrained on some large scale, because
the link to the ``backreaction term'' makes sure that, at any time, the ``interior'' of
a universe model (below this scale) also determines the global curvature. In other 
words: there is no analogue of the {\em Construction Principle}.
This comment is not entirely correct, since further study may reveal that it could be
possible to formulate an analoguous principle on the basis of topological constraints
that may be imposed on the space sections. In any case, research should not lack 
commitment to the
challenge of constructing inhomogeneous evolution models on curved space sections,
since this is evidently the generic case. We note that existing work on 
general relativistic evolution models and their averages relies in most cases on
assumptions that are closely designed after the Newtonian case in order to 
rescue standard procedures like periodic boundary conditions, decomposition of
inhomogeneities into plane waves and set--ups of initial conditions on locally flat
space sections (compare {}\cite{russ:backreaction} and comments in 
{}\cite{buchert:onaverage} and {}\cite{buchert:dialogue}). 

\subsection{What Einstein Wanted}

In the previous subsection we advocated a viewpoint that concentrates on the 
physics of fluids, which (actively) determines the geometry of spacetime, 
in particular its average properties. As a showcase we learned in the 
Newtonian framework that
the morphological properties of spatial domains are determined by the 
averages over fluctuations in the kinematical variables (Subsection 1.12). 
A similar view may also apply to the relativistic context and, most interestingly,
also to the global morphology of space sections.

Turning this around, we may alternatively take the (passive) viewpoint of  
understanding the global world structure as given in terms of 
geometrical (and topological) conjectures (which may themselves be subjected to
observational falsification; an example is the possibility of falsifying 
possible topological spaceforms on the basis of microwave background observations
({}\cite{cornish:circles}; see also {}\cite{uzan:topology} and ref. therein).

Einstein's vision of a globally static universe model stands out as a famous example.
His ``prejudice'' that the Universe as a whole should be static and all evolution
(assigning sense to the notion of time) should take place in the ``interior''
of this world model entails a strong determination of the model's average properties.
Only his invention of the cosmological term made this vision possible resulting in
the beautiful structure of a spherical space in which, apart from topological
degeneracies, many characteristics (like the radius of the universe model) are 
practically fixed.
Hence, according to what we have learned about average models, the fluid's
fluctuations are  ``slaved'' to the global predetermined structure. This global 
model even guarantees strong evolution, because   
fluctuations evolve exponentially in a static cosmos.

The problem that his model is unstable (within the class of FLRW models) applies also
to the other expanding candidates, if the class is widened to include inhomogeneities.
Einstein's example so serves as a guide to think about the average model we have
discussed in a geometrical way. The assumption of a static model
(the general expansion law allows for a static cosmos even without the 
cosmological term) should be taken as an example for the possible 
slaving of fluctuations to global structural assumptions. Of course, other 
such assumptions are possible. This illustrates how  global geometrical 
and topological constraints on
spaceforms could provide ``boundary conditions'' for the evolution of fluctuations.
Since, in Newtonian cosmology, we are always working on a toroidal space without 
further questioning this assumption, this point of view is not exotic, but 
plants the seed for possibly fruitful research directions. 
That these directions are mathematically very involved can be made obvious by 
pointing out the relation between the average curvature and its compatibility with
topological spaceforms (a spaceform with globally negative spatial curvature
is not compatible with simply--connected space sections).
Due to the process of structure formation the boundary of the domain may break, or
it may self--intersect and split into two domains (these events are
topologically classified in the framework of Legendrian singularities, 
{}\cite{arnold:singularities} and ref. therein).
Since the spatial metric is linked to the fluid, a singular dynamics could 
induce topology changes. 
Contrary to the existence of such natural
metamorphoses, the averager we have defined relies on the assumption that the domain 
remains simply--connected (Fig.~\ref{fig:jgrg_buchert3}).
 
\subsection{Expansion Law of General Relativity: Perfect Fluid Models}

As an outlook I briefly give one of the results obtained by averaging a perfect
fluid cosmology on hypersurfaces $S(t)=const.$ as explained above.
  
The averaged equations can be summarized in the following way {}\cite{buchert:grgfluid}:

Let us define effective densities as sources of an expansion law,
\begin{equation}
\label{eq:effective2}
\varepsilon_{\rm eff}:= \average{\tilde\varepsilon} - 
\frac{{\tilde{\cal Q}}_{\CD}}{16\pi G} \;\;\;,\;\;\;
p_{\rm eff}:= \average{\tilde p} - 
\frac{{\tilde{\cal Q}_{\CD}}}{16\pi G} 
- \frac{{\tilde{\cal P}}_{\CD}}{12\pi G}\;\;\;,
\end{equation} 
with the scaled matter sources ${\tilde\varepsilon} :=
N^2 \varepsilon$ and ${\tilde p} :=N^2 p$. Then, the averaged equations 
can be cast into a form similar to the standard Friedmann equations:

\begin{equation}
\label{eq:effectiveform2a}
3\frac{{\partial_t^2 a}_{\CD}}{a_{\CD}} + 4\pi G 
\left(\varepsilon_{\rm eff} 
+ 3p_{\rm eff}\right)\;=\; 0\;\;\;;
\end{equation}

\begin{equation}
\label{eq:effectiveform2b}
6 {\tilde H}_{\CD}^2 + \langle{\tilde{\cal R}}\rangle_{\CD} 
- 16\pi G \varepsilon_{\rm eff} 
\;=\; 0\;\;\;,
\end{equation}
and the integrability condition of Eq.~(\ref{eq:effectiveform2a}) 
to yield Eq.~(\ref{eq:effectiveform2b}) has the form of a 
balance equation between the effective sources and the averaged spatial 
(t--)Ricci scalar ${\tilde{\cal R}} := N^2 {\cal R}$:

\begin{equation}
\label{eq:effectiveintegrability2}
\partial_t \varepsilon_{\rm eff} + 3 {\tilde H}_{\CD}
\left(\varepsilon_{\rm eff} + p_{\rm eff}\right)
 = {1\over 16\pi G} \left( \partial_t \langle{\tilde{\cal R}}\rangle_{\CD} +
 2 {\tilde H}_{\CD} \langle{\tilde{\cal R}}\rangle_{\CD} \right)\;\;\;.
\end{equation}

The effective densities obey a conservation law, if the domains' curvature
evolves like in a ``small'' FLRW cosmology, 
$\langle{\tilde{\cal R}}\rangle_{\CD} =0$, or 
$\langle{\tilde{\cal R}}\rangle_{\CD} \propto a_{\CD}^{-2}$, respectively.

The expressions ${\tilde{\cal Q}}_{\CD}$ and ${\tilde{\cal P}}_{\CD}$ are given in
{}\cite{buchert:grgfluid}. As can be seen above, ${\tilde{\cal Q}}_{\CD}$ (the 
``kinematical backreaction'') acts like a fluid component with ``stiff'' equation of
state, similar to the action of a minimially coupled scalar field, while
${\tilde{\cal P}}_{\CD}$ (the ``dynamical backreaction'') is due to the
non--vanishing pressure gradient in the hypersurfaces. 

It is interesting that the ``kinematical backreaction'' of the inhomogeneities 
could play the role of a free scalar field component that is commonly introduced
when early evolutionary stages of the Universe are studied. 
It is also interesting that the expansion law in the case of a minimally coupled
scalar field source gets particularly simple. This case will be investigated in a 
forthcoming work {}\cite{buchert:grgscalarfield}.   

\section*{Acknowledgements}

I would like to thank the organizers of this workshop for their invitation and
support during a pleasant and 
inspiring stay at Hiroshima University. I also appreciate generous support
and hospitality by the  National Astronomical Observatory in Tokyo, as
well  as  hospitality  at  Tohoku  University in  Sendai.  
Also, I  would like to thank my collaborators with whome I share some of the
reported results.

\end{document}